\documentclass[12pt]{iopart}

\usepackage{graphicx}
\usepackage{epic,eepic}
\usepackage{graphics}
\usepackage{epsfig}
\usepackage{subfigure}

\newcommand{\be}{\begin{equation}}
\newcommand{\ee}{\end{equation}}
\newcommand{\bea}{\begin{eqnarray}}
\newcommand{\eea}{\end{eqnarray}}

\newcommand{\T}{T}

\newcommand{\dt}{{\Delta t}}
 
\newcommand{\pii}{{\partial_i}}

\newcommand{\pj}{{\partial_j}}

\usepackage{amsfonts}
\usepackage{amssymb}
\usepackage{bm}
\usepackage{color}
\usepackage{graphicx}

\begin{document}
\title{Numerical simulations of compressible Rayleigh-Taylor turbulence in stratified fluids.}

\author{A. Scagliarini} \address{Department of Physics and INFN,
  University of Tor Vergata,\\Via della Ricerca Scientifica 1, 00133
  Rome, Italy\\and International Collaboration for
  Turbulence Research} 
\author{L. Biferale} \address{Department of Physics and INFN,
  University of Tor Vergata,\\Via della Ricerca Scientifica 1, 00133
  Rome, Italy\\and International Collaboration for
  Turbulence Research} 
\author{M. Sbragaglia} \address{Department of Physics and INFN, University of Tor
  Vergata, \\ Via della Ricerca Scientifica 1, 00133 Rome, Italy}
\author{K. Sugiyama} \address{Department of Mechanical Engineering, School of Engineering,\\ The University of Tokyo, 7-3-1, Hongo, Bunkyo-ku, Tokyo 113-8656, Japan} \author{F. Toschi} \address{Department of Physics and Department of Mathematics and
  Computer Science, Eindhoven University of Technology, 5600 MB
  Eindhoven, The Netherlands; and International Collaboration for
  Turbulence Research}

\begin{abstract}
We present results from numerical simulations of Rayleigh-Taylor turbulence, performed using a  recently proposed \cite{JFM,POF} lattice Boltzmann method able to describe consistently a thermal compressible flow subject to an external forcing. The method allowed us to study the system both in the nearly-Boussinesq and strongly compressible regimes. Moreover, we show that when the stratification is important,  the presence of the adiabatic gradient causes the arrest of the mixing process. 
\end{abstract}

\maketitle

\section{Introduction}
Computational methods based on discrete-velocity models have
gained considerable interest in the recent period as efficient tools
for the theoretical investigation of the properties of complex flows \cite{SC1,Yeo,HeLuo,Ladd,ladd-review}. In particular, it has been recently shown that an important class of these models, known as the lattice Boltzmann models (LBM) \cite{Gladrow,BSV,Chen}, can be derived from the continuum Boltzmann (BGK) equation \cite{BGK54}. This derivation involves the expansion in suitable Hermite polynomials of the distribution functions $f({\bm x}, {\bm \xi},t)$, describing the probability of finding a molecule at space-time location $({\bm x},t)$ and with velocity ${\bm  \xi}$ \cite{HeLuo,ShanHe98,Martys98,Shan06}. Therefore, the corresponding lattice dynamics is well founded in terms of an underlying continuum kinetic theory. The state-of-the-art is satisfactory concerning iso-thermal flows, even in presence of complex bulk physics (multi-phase, multi-components)
\cite{SC1,Yeo,HeDoolen01} and/or with complex boundary properties
 including roughness, wetting and slip boundary conditions \cite{pre.nostro,prl.nostro,Harting}.\\
The situation is much less satisfactory when hydrodynamical temperature fluctuations play an active role in the flow evolution, due to complex compressible effects or to phase-transition in multi-phase systems. 

Within this framework, we have recently developed \cite{JFM,POF} a new LBM which allows to incorporate the effects of external/internal forces in  thermal systems. Here we  use this new algorithm to study highly compressible Rayleigh-Taylor systems, with an initial configuration such that two blobs of the same fluid are prepared with two different temperatures (hot, less dense, blob below, cold, denser, blob above). We show that the method is able to handle the highly non-trivial spatio-temporal evolution of the system even in the developing turbulent phase. In this case, we could push the numerics up to Atwood numbers $At \sim 0.4$. Maximum Rayleigh numbers achieved are $Ra \sim 4\times 10^{10}$ for $At=0.05$ and $Ra \sim 2\times 10^{9}$ for $At=0.4$.  The paper is organized as follows. We will first describe the method (section \ref{sec:LBM}),  the numerical setup (section \ref{sec:sim}) and the system studied (section \ref{sec:RT}); then we will discuss the two main physical results, i.e. the stratification (section \ref{sec:adiagra}) and  compressibility (section \ref{sec:pdfs}) effects, and some features related to
the conservation of mean quantities (section \ref{sec:conservation}). Conclusions are in  section  \ref{sec:conclusions}.

\section{The Lattice Boltzmann Model}\label{sec:LBM}
Here we introduce the simulated equations set along with a brief description of the computational lattice Boltzmann method employed. More details, along with many validations, can be found in \cite{POF}.
The thermal-kinetic description of a compressible gas/fluid of variable density $\rho$, local velocity ${\bm u }$, internal energy ${\cal K}$, and subject to a local body force density ${\bm g}$, is given by the following equations: 
\begin{eqnarray} \label{eq:conslaws}
\partial_t \rho + \partial_i (\rho u_i) = 0 \nonumber \\
\partial_t (\rho u_k) + \partial_i (P_{ik}) = \rho g_k \nonumber \\ 
\partial_t {\cal K} + \frac{1}{2} \partial_i q_i = \rho g_i u_i 
\end{eqnarray}
where $P_{ik}$ and $q_i$ are the momentum and energy fluxes, describing advection, viscous properties and thermal diffusivities in the hydrodynamical limit.

 In \cite{JFM} it has been shown that it is possible to recover exactly these equations, starting from a continuum Boltzmann Equation and introducing a suitable shift of the velocity and temperature fields entering in the local equilibrium.

\begin{figure}
\begin{center}
  \includegraphics[scale=0.3]{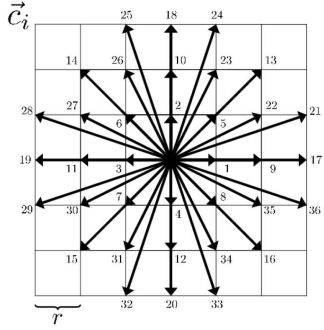}
  \caption{Scheme of the discrete set of velocities; $r$ is the lattice constant whose value is $r=1.1969...$ 
 (see \cite{POF} and references therein).}
\label{fig:37}
\end{center}
\end{figure}

The lattice counterpart of the continuum description can be obtained through the lattice Boltzmann discretization ($f_l(x,t)$ are the fields associated to the populations):
\be \label{eq:LBM}
f_{l}({\bm x}+ {\bm c}_{l} \dt,t+\dt) - f_{l}({\bm x},t)=-\frac{\dt}{\tau}\left(f_{l}({\bm x},t) - f_l^{(eq)}\right) 
\ee
where the equilibrium ${f}_{l}^{(eq)}$ is expressed in terms of hydrodynamical fields and the  body force term ${\bm g}$, and the subscript $l$ runs over 
the discrete set of velocities ${\bm c}_l$ (see fig.~\ref{fig:37}); in equation (\ref{eq:LBM}) $\tau$ is the relaxation time
(which is related to the dynamic viscosity $\nu$ via $\nu = \rho T (\tau - 1/2)$, $T$ being the temperature field), and $\Delta t$ the time step of the simulation.
  
The macroscopic fields (density, momentum and temperature) are defined in terms of the lattice Boltzmann populations: 
$\rho = \sum_l f_l$, $\rho {\bm u} = \sum_l {\bm c}_l f_l$, $D \rho \T = \sum_l \left|{\bm c}_l - {\bm u}\right|^2  f_l$.
Lattice discretization induces non trivial corrections terms in the macroscopic evolution of averaged hydrodynamical quantities. In particular, both momentum and temperature must be renormalized by discretization effects in order to recover the correct description out of the discretized LBM variables: the first correction to momentum is given by the pre and post-collisional average
\cite{Buick00,Guo02} $\label{eq:shift_lattice} {\bm u}^{(H)}={\bm u}+\frac{\Delta t}{2} {\bm g}$ and the first, non-trivial, correction to the temperature field by \cite{JFM} $ {\T}^{(H)} = \T + \frac{(\dt)^2g^2}{4 D }. \label{eq:thydro}$ ($D$ is the dimensionality of the system). Using these ``renormalized'' hydrodynamical fields it is possible to recover, through a Taylor expansions in $\Delta t$, the thermo-hydrodynamical equations \cite{JFM,POF}:
\bea\label{eq:H1}
D_t \rho = - \rho \pii u^{(H)}_i  \\ \label{eq:H2}
\rho D_t u^{(H)}_i = - \pii p - \rho g\delta_{i,3} + \nu \partial_{jj}
u^{(H)}_i  \\\label{eq:H3} 
\rho c_v D_t T^{(H)} + p \pii u^{(H)}_i = k \partial_{ii} T^{(H)} 
\eea
where we have introduced the material derivative, $D_t = \partial_t + u_j^{(H)} \pj$, and we have neglected viscous dissipation in the temperature 
equation (usually small). Moreover, $c_v$ is the specific heat at constant volume  for an ideal gas $p=\rho T^{(H)}$, and $\nu$ and $k$ are the 
transport coefficients. From now on, for the sake of simplicity, we will drop the superscript $(H)$, knowing that we are dealing with lattice
hydrodynamical quantities satisfying equations  (\ref{eq:H1}-\ref{eq:H3}).
As a tool for the numerical simulation of systems as (or similar to) the one we plan to study, the LBM may suffer, in principle,
of some issues, such as having too high Mach numbers, too low viscosity (i.e. very small relaxation time, which is undesirable especially
in the presence of processes very far from local equilibrium): we could, however, check the accuracy of our method against different ones,
finding it extremely competitive, within the range of parameters discussed in this paper \cite{POF2}.

\section{Details of the numerical simulations} \label{sec:sim}
We use a 2D LBM algorithm, with 37 population fields
(a so called D2Q37 model), moving with the lattice speeds shown in figure (\ref{fig:37}). 
Since the lattice spacing can be taken to be unitary, the time step $\Delta t$ will be the inverse
of the lattice unit speed, i.e. $\Delta t \sim 0.835...$.
Three different kinds of simulations have been performed (whose parameters are summarized in table \ref{table:param}):
(A) with a large enough adiabatic gradient (but small Atwood number) in order to address the 
stratification effects on the mixing layer growth, while still being very close to the Boussinesq approximation;
(B) with an adiabatic gradient which is twice the one of run A;
(C)  with large Atwood in order to describe compressibility effects, out of the Boussinesq regime, but 
far from the adiabatic profile;
(D)  with small adiabatic gradient and small Atwood number.
\begin{table*}
\begin{center}
\begin{tabular}{|c | c c c c c c c c |}
  \hline & $At$ & $L_x$ & $L_z$ & $\nu$ & $g$ & $T_u$ & $T_d$ & $\tilde{\tau}$ \\
\hline run A & $0.05$ & 800 & 1400 & 0.001 & $2.5 \times 10^{-4}$ & $0.95$ & $1.05$ & $8 \times 10^3$ \\
run B &$0.05$& 800 & 1400 &0.001 & $5 \times 10^{-4} $ &
$0.95$ & $1.05$ & $ 5.6 \times 10^3$ \\ 
run C &$0.4$& 1664 & 4400 &0.1 & $1 \times 10^{-4} $ &
$0.6$ & $1.4$ & $ 6.5 \times 10^3$ \\ 
run D &$0.05$& 1024 & 2400 &0.005 & $5 \times 10^{-5} $ &
$0.05$ & $1.05$ & $ 2 \times 10^4$ \\ \hline
\end{tabular}
\caption{Parameters for the three types of Rayleigh-Taylor runs. Atwood number, $At=(T_d-T_u)/(T_d+T_u)$;
viscosity $\nu$; gravity $g$; temperature in the upper half region, $T_u$; temperature in the lower half region, $T_d$;
normalization time, $\tilde{\tau}=\sqrt{L_x/(g\;At)}$.}
\label{table:param} 
\end{center}
\end{table*}

\section{The Rayleigh-Taylor system} \label{sec:RT}
\begin{figure}
  \begin{center}
    \advance\leftskip-2.9cm
    \includegraphics[width=0.4\textwidth]{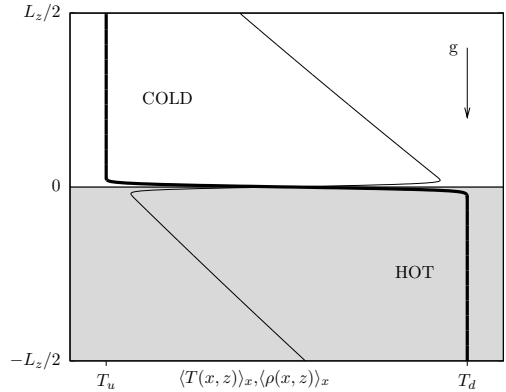}
    \caption{Initial condition of the Rayleigh-Taylor system, as given by equation (\ref{eq:inithydro}). We show
    the mean temperature (bold line) and density (tiny line) profiles as function of the $z$-coordinate (on the vertical axis).}
    \label{fig:rt}
  \end{center}
\end{figure}

Superposition of a heavy fluid above a lighter one in a constant
acceleration field depicts a hydrodynamic unstable configuration
called the Rayleigh-Taylor (RT) instability \cite{chandra} with applications on
different fields going from inertial-confinement fusion \cite{rt2} to
supernovae explosions \cite{rt3} and many others \cite{rt.review}.
Although this instability was studied for decades it is still an open
problem in several aspects \cite{rt4}. In particular, it is crucial to
control the initial and late evolution of the mixing layer between the
two miscible fluids; the small-scale turbulent fluctuations, their
anisotropic/isotropic ratio; their dependency on the initial
perturbation spectrum or on the physical dimensions of the embedding
space \cite{jot,boffi}. In many cases, especially concerning
astrophysical and nuclear applications, the two fluids evolve with
strong compressible and/or stratification effects, a situation which
is difficult to investigate either theoretically or numerically. Here,
we concentrate on the large scale properties of the mixing layer,
studying a slightly different RT system than what usually found in the
literature: the spatio temporal evolution of a single component fluid
when initially prepared on the hydrostatic unstable equilibrium,
i.e. with a cold uniform region in the top half and a hot uniform
region on the bottom half (see figure \ref{fig:rt}).
For the sake of simplicity we limit the investigation to the 2d
case. While small-scales fluctuations may be strongly different in 2d
or 3d geometries, the large scale mixing layer growth is not supposed
to change its qualitative evolution \cite{chertkov,cabot}.
\begin{figure}
  \begin{center}
    \advance\leftskip-2.9cm
    \includegraphics[width=0.8\textwidth]{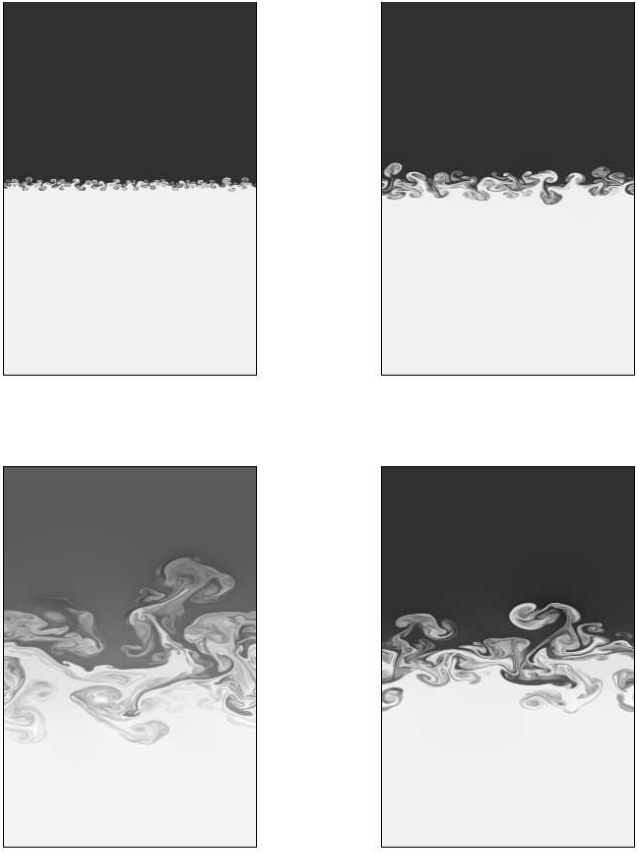}
    \caption{Spatio-temporal evolution for a typical RT run with
      $L_x\times L_z = 1024 \times 2400$, $T_u=0.95$, $T_d=1.05$ at four instant of 
      time: $t =  \tilde{\tau}, 2\tilde{\tau}, 4\tilde{\tau}, 6\tilde{\tau}$ (run D in table \ref{table:param}) 
     going clockwise from the top left panel.}
    \label{fig:rt.evolution}
  \end{center}
\end{figure}
Grey-scale coded snapshots of a typical RT  evolution are shown in
figure \ref{fig:rt.evolution} showing all the complexity of the
phenomena.  Let us start to define precisely the initial set-up. We
prepare a single component compressible flow in a 2d tank of size,
$L_x \times L_z$, with adiabatic and no-slip boundary conditions on
the top and bottom walls, and with periodic boundary conditions on the
vertical boundaries. For convenience we define the initial interface
to be at height $z=0$, the box extending up to $z=L_z/2$ above and
$z=-L_z/2$ below it (see figure \ref{fig:rt}). In the two half volumes
we then fix two different homogeneous temperature, with the
corresponding hydrostatic density profiles, $\rho_0$, verifying
\be
\label{eq:phydroRT}
\partial_z p_0(z) = -g \rho_0(z).  
\ee 
Considering that in each half we have $p_0(z) =T \rho_0(z)$,with $T$ fixed, the solution has an
exponentially decaying behavior in the two half volumes, each one
driven by its own temperature value.  The initial hydrostatic unstable
configuration is therefore given by:
\begin{eqnarray} \label{eq:inithydro}
  T_0(z) = T_{u};\,\, \rho_0(z) = \rho_u \exp(-g(z-z_c)/T_u);\qquad z >0 \nonumber \\
  T_0(z) = T_{d};\,\, \rho_0(z) = \rho_b \exp(-g(z-z_c)/T_d);\qquad z <0. \nonumber \\
\end{eqnarray}
To be at equilibrium, we require to have the same pressure at the
interface, $z=z_c=0$; which translates in a simple condition on the
prefactor of the above expressions: \be \rho_u T_u = \rho_b T_d.  \ee
Because $T_u<T_d$, we have at the interface $\rho_u> \rho_b$.  As far
as we know, there are no exhaustive detailed calculations of the
stability problem for such configuration, even though not too
different from the usual RT compressible case
\cite{chandra,gauthier3,gauthier.priv}.  As said, this is not the
common way to study RT systems, which is usually meant as the
superposition of two different miscible fluids, isothermal, with
different densities \cite{chandra,gauthier2,gauthier3,jot}. As far as
compressible effects are small, one may safely neglect pressure
fluctuations and write -- for the case of an ideal gas: 
\be \label{eq:fluctrho}
\frac{\delta \rho}{\rho} \sim -\frac{\delta T}{T} 
\ee 
and the two RT
experiments are then strictly equivalent. Moreover, in the latter
case, if one may neglect the dependency of viscosity and thermal
diffusivity from temperature, the final evolution is undistinguishable
from the evolution of the temperature in the Boussinesq approximation
\cite{chertkov,boffi}.

\section{The adiabatic gradient and the arrest of the mixing process} \label{sec:adiagra}
\begin{figure}
\begin{center}
\advance\leftskip-0.55cm
\includegraphics[width=0.5\textwidth]{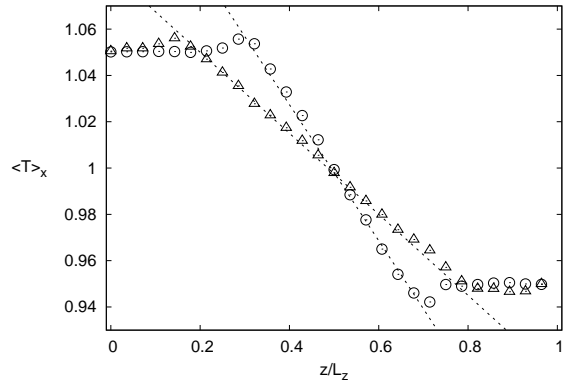}
\caption{Mean temperature profiles $\langle T \rangle_x(z,t)$ for run A ($\triangle$) and run B ($\odot$), in a stage where the mixing 
process is already stopped ($t \simeq 13 \tilde{\tau}$ for both cases). The dashed lines represent the correspoding adiabatic profiles.}
\label{fig:tempprofs}
\end{center}
\end{figure}

The main novelty in the system here investigated is due to the
presence of new effects induced by the adiabatic gradient,
which can be written for an ideal gas as
$\beta_{ad} = g/c_p$. 
The role of stratification, i.e. of the adiabatic gradient, is quite well established in the context of Rayleigh-B\'enard convection
(see, e.g. \cite{tritton}), while it has been only in recent times studied, numerically \cite{rast,milovich} and theoretically \cite{abarzhi}, 
in a set up such as that of Rayleigh-Taylor mixing.  
In order to understand the main physical point
it is useful to think at the RT mixing layer as equivalent to a
(developing) Rayleigh-B\'enard system with an imposed mean temperature
gradient \cite{celani1,celani2}.  Let us denote with $L_{ml}(t)$ the typical
width of the RT mixing layer at a given time as measured for example
from the distance between the two elevations where the mean
temperature profile is $1\%$ lower or higher then the bottom and top,
respectively, unmixed temperature values, $L_{ml} = z_u-z_d$, where
$\langle T (x,z_{u}) \rangle_x = 1.01 T_u$ and $\langle T (x,z_{d})                                                      
\rangle_x = 0.99 T_d$.  The temperature tends to
develop a linear profile inside the mixing region, the resulting
instantaneous temperature gradient being given by $\beta(t) =                                                          
(T_d-T_u)/L_{ml}(t)$, and, hence, decreasing in time inversely to the growth
of the mixing length.  As a result, at a certain time (if the box is high enough) 
the instantaneous temperature gradient will become of the same
order of the adiabatic gradient, $\beta(t) \sim \beta_{ad}$ and the
growth of the mixing length will stop. 
In figure (\ref{fig:tempprofs}) we show the mean temperature profiles
once the mixing is already stopped, for two different values of gravity (runs A and B in table (\ref{table:param})): in
the mixing layer the two curves have developed a linear profile with slope $g/c_p$, which is exactly the adiabatic gradient for an ideal gas.
One can define an instantaneous
Rayleigh number, driving the physics inside the mixing layer, as:
\be
\label{eq:Rat}
\tilde Ra(t) = \frac{(g/\tilde T_0) L_{ml}^4(t)
  (\beta(t)-\beta_{ad})}{(k/ \tilde \rho_0 c_p)(\nu/ \tilde \rho_0)},
\ee
where $\tilde{ (\cdot)}$ indicates quantities evaluated at the
middle layer. It is clear that for small times, $\beta(t) \ll                                                            
\beta_{ad}$, the effective instantaneous Rayleigh number is high: the
system is unstable, and the mixing length grows. On the other hand, as
time elapses, the vertical mean temperature gradient decreases, until
a point when, $\beta(t)\sim \beta_{ad}$, the instantaneous effective
Rayleigh number becomes $\tilde Ra(t) \sim {\cal O}(1)$ and the system
tends to be stabilized. We can then identify an {\it adiabatic length}: 
$$
L_{ad} = (T_d-T_u)/\beta_{ad} = c_p \Delta T/g
$$ 
which determines the maximum length achievable by the mixing layer, in our
configuration. 
When the mean temperature approaches the adiabatic profile, the system shows a
sudden slowing down of the mixing layer growth which, eventually, stops. 
A possible way to estimate quantitatively when and how the adiabatic
gradient starts to play a role in the growth of the mixing length is
to use a simple phenomenological closure for large scale quantities in
the system.  We start from the self-similar scaling predicted by
\cite{cook,cook2} for the homogeneous unstratified growth:
\be
\label{eq:cook}
(\dot L_{ml}(t))^2 = 4 \alpha\, g\, At\, L_{ml}(t)
\ee
which has a unique solution  in terms of the initial value, $L_{ml}(t_0)$:
\be
\label{eq:Lt5}
L_{ml}(t) = L_{ml}(t_0) +  2\,\sqrt{\alpha\, At\, g}\,  (t-t_0)  + \alpha\, At\,g\, (t-t_0)^2.
\ee
In order to minimally modify the above  expression considering the saturation effects induced by stratification, 
we proposed to use in \cite{POF}:
\be
\label{eq:luca-andrea}
(\dot L_{ml}(t))^2 = 4 \alpha\, g\, At\, L_{ml}(t)\psi\left(\frac{L_{ml}(t)}{L_{ad}}\right)
\ee
where $\psi=\psi(x)$ must be a function fulfilling the condition $\psi \rightarrow 1$ as $ x \rightarrow 0$
(that is for $L_{ad} \rightarrow \infty$), in order to recover the equation (\ref{eq:cook}) for the unstratified case
when the adiabatic gradient goes to zero. We further add the
requirement of reaching the adiabatic profile with
zero velocity and acceleration, enforcing a strict irreversible growth, i.e. $\dot L_{ml} \ge 0$, as it must
be for the case of miscible fluids. Under these assumptions it can be shown that the simplest form 
for the function $\psi$ is:
\be
\label{eq:luca-andrea1}
\psi\left(\frac{L}{L_{ad}}\right) = C\left[e^{-\left(\frac{L-L_{ad}}{L_{ad}}\right)}-
\left(\frac{2L_{ad}-L}{L_{ad}}\right)\right]
\ee
where the prefactor $C$ must be set equal to $1/(e-2)$ to comply with the prescribed boundary conditions.
Equation (\ref{eq:luca-andrea}) must be considered as a zero-th order phenomenological way to take into account 
for the adiabatic gradient in the mixing layer evolution.
\begin{figure}
\begin{center}
\advance\leftskip-0.55cm
\includegraphics[width=0.5\textwidth]{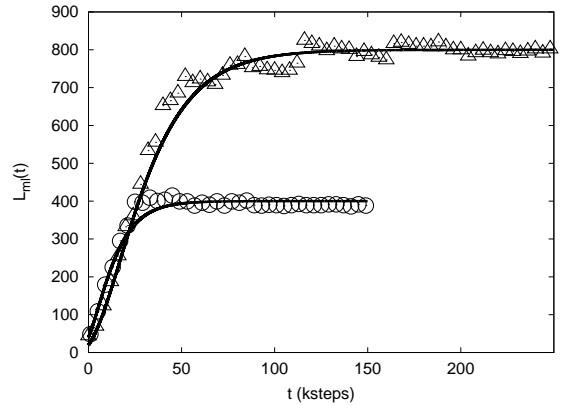}
\caption{Evolutions of the mixing layer, $L_{ml}(t)$ versus time with two different adiabatic lengths, corresponding
to run A ($\triangle$) and run B ($\odot$) in table \ref{table:param}.
Solid bold lines correspond to the theoretical prediction (\ref{eq:luca-andrea}) with $\alpha                                                         
=0.05$.}
\label{fig:adiabatic.profile}
\end{center}
\end{figure}
We integrated numerically eq. (\ref{eq:luca-andrea}) testing the
result in figure \ref{fig:adiabatic.profile} where we show that
it is possible to fit the global evolution of the mixing length
$L_{ml}(t)$, by using reasonable \cite{rt4} values of $\alpha$, for all times,
including the long time behavior where $L_{ml}(t) \sim L_{ad}$. 
We can then interpret the
solution of our equation (\ref{eq:luca-andrea}) as a good
generalization of (\ref{eq:Lt5}) including also the adiabatic
gradient effects.
\section{Effects of compressibility} \label{sec:pdfs}
\begin{figure}
\begin{center}
\advance\leftskip-0.55cm
\includegraphics[width=0.5\textwidth]{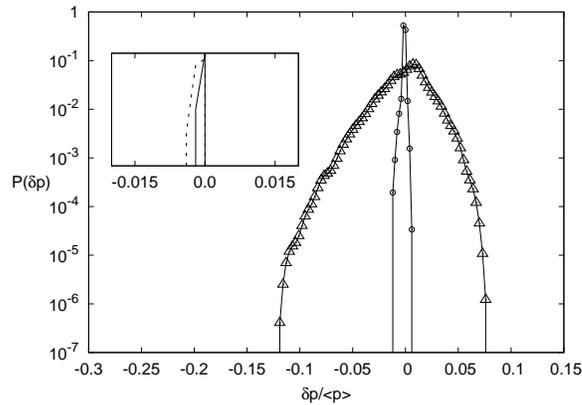}
\caption{Probability density functions of pressure fluctuations at time $t=\tilde{\tau}$ ($\odot$) and 
$t=5\tilde{\tau}$ ($\triangle$) for $At=0.4$ and (inset) for $At=0.05$ ($t=\tilde{\tau}$, solid line, and $t=5\tilde{\tau}$, dashed line): 
while in the low $At$ case the PDF remains basically a $\delta-$function at any time,
it is more spread (with tails becoming larger as the time increases) in the compressible case ($At=0.4$).}
\label{fig:pdfP.DT04+DT005}
\end{center}
\end{figure}

In this section we are going to study the effects of flow compressibility 
on the dynamics of a RT system at changing  Atwood number.
To do that we come back to the discussion sketched in section \ref{sec:RT}; since the equation
of state for our fluid is the one of a perfect gas, the pressure, temperature and density 
fluctuations are related by:
\be
\frac{\delta P}{P} = \frac{\delta \rho}{\rho} + \frac{\delta T}{T},
\ee
hence, as discussed in section \ref{sec:RT}, if pressure fluctuations are small, density
fluctuations are linearly slaved on those of temperature, which
will be also small, and the 
system behaves as a Boussinesq fluid. Conversely, if the temperature jump is high there will
be large density differences through the system (\ref{eq:inithydro}) and hence 
large pressure fluctuations (\ref{eq:phydroRT}); thus we expect that, at increasing the 
Atwood number, the dynamics becomes more and more compressible and pressure turns out to be a dynamically
relevant variable. 
\begin{figure}
\begin{center}
\advance\leftskip-0.55cm
\includegraphics[width=0.5\textwidth]{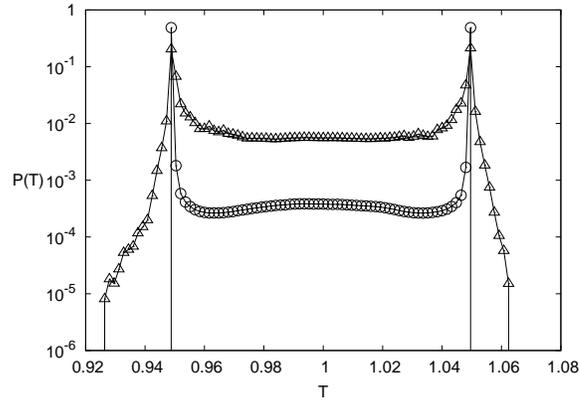}
\caption{Probability density functions of temperature at $t=\tilde{\tau}$ ($\odot$) and at $t=5\tilde{\tau}$ ($\triangle$), for $At=0.05$, for which the flow is basically Boussinesq-like.}
\label{fig:pdfT.DT005}
\end{center}
\end{figure}
\begin{figure}
\begin{center}
\advance\leftskip-0.55cm
\includegraphics[width=0.5\textwidth]{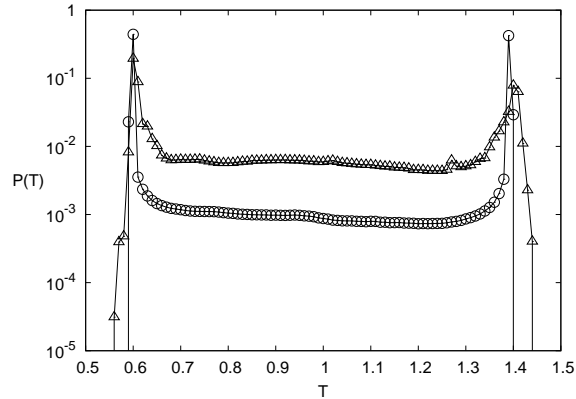}
\caption{Probability density functions of temperature at $t=\tilde{\tau}$ ($\odot$) and at $t=5\tilde{\tau}$ ($\triangle$), for $At=0.4$, that is in the
highly compressible regime.}
\label{fig:pdfT.DT04}
\end{center}
\end{figure}
We show, first, how the mixing acts on the statistics of the (point-wise) temperature; as we can see in figures
(\ref{fig:pdfT.DT005}) and (\ref{fig:pdfT.DT04}), 
where we plot the probability density function (PDF) of temperature at two instants of time for the 
two $At=0.05$ and $At=0.4$ (for which the squared Mach number is $Ma^2 \sim 0.16$), 
the distribution has initially a bimodal character, since at the beginning the 
volume is divided into two homogeneous regions of hot and cold fluid. Due to the mixing, at later times, the probability
of having intermediate values of temperature increases; however, the two peaks remain dominant, because the system
dynamics does not involve yet the whole box and because the diffusive processes are so slow to be irrelevant at this
stage. No evident differences (except the obviously larger range of values spanned) rise between the low and high 
Atwood number cases. Hence, to better address this point, we study the statistics of pressure fluctuations
(with respect to the mean profile), in the two compressibility regimes; we define the fluctuation
of the generic thermohydrodynamic field $\phi$ as 
$$
\delta \phi (\mathbf{x},t) = \phi (\mathbf{x},t) - \langle \phi \rangle_x (z,t); \quad 
\langle \phi \rangle_x (z,t) = \int_0^{L_x} \phi \; dx.
$$
In figure (\ref{fig:pdfP.DT04+DT005}) we show the PDFs of $\delta p$, 
measured again inside the whole volume, at 
two different instants of time during the mixing process, for two Atwood numbers. 
It can be noticed how the PDF, while being basically a $\delta$-function
for low $At$ and remaining such at any time (see inset of figure (\ref{fig:pdfP.DT04+DT005})), 
is more spread at higher $At$ and enlarges its tails as the time elapses, 
confirming that pressure dynamics is now highly non-trivial.
\\

It is, moreover, known that increasing the degree of compressibility of the dynamics has also a strong impact 
in the stability properties of the system
\cite{livescu} and in the statistics of the mixing layer growth process, determining in the latter case an asymmetry
in the growth of the mixing layer \cite{livescu}, noticeable also in the statistics of the growth parameter $\alpha$ \cite{POF}.  
We would like to discuss here such effects, without appealing to any phenomenological model, but in terms, again, of
PDFs of the temperature field. 
The use of PDFs to address compressibility effects in RT was also suggested, although
in a slightly different way, in \cite{glimm}, in regimes from low ($Ma^2 \sim 0.008$)
to moderately high ($Ma^2 \sim 0.1$) squared Mach number.
With this aim, we measured the $\mathcal{P}(T)$ where 
$T=T(x,z^{\ast},t^{\ast}=5\tilde{\tau})$ along lines at two fixed heights $z^{\ast}$ 
(at a certain time in the late stage of the evolution), 
within the mixing layer, symmetrically displaced with respect to the mid cell;
in particular we chose $z^{\ast} = \pm L_{ml}(t)/2$.  
\begin{figure}
\begin{center}
\advance\leftskip-0.55cm
\includegraphics[width=0.5\textwidth]{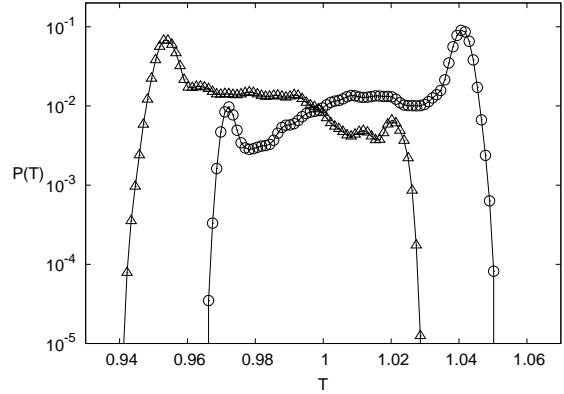}
\caption{PDFs of the temperature field $T(x,z^{\ast},t=5\tilde{\tau})$, measured at $z^{\ast}=\pm L_{ml}(t)/2$, for $At=0.05$. The two PDFs show peaks close
to the values of $T$ of the unmixed cold (at $z^{\ast} = +L_{ml}/2$, ($\triangle$)) and hot (at $z^{\ast} = -L_{ml}/2$, ($\odot$)) fluid respectively; the two PDFs are symmetric to each other with respect to the mean temperature value $T=1$ (typical phenomenology for a Boussinesq system).}
\label{fig:pdfTz.DT005}
\end{center}
\end{figure}
\begin{figure}
\begin{center}
\advance\leftskip-0.55cm
\includegraphics[width=0.5\textwidth]{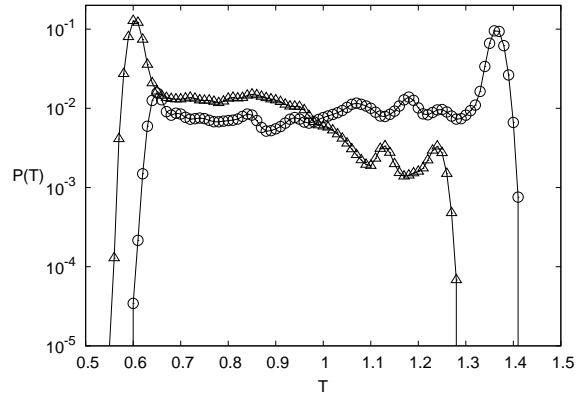}
\caption{Same as figure (\ref{fig:pdfTz.DT005}) but for $At=0.4$. 
Differently from the Boussinesq (low $At$) case, the two PDFs are not symmetric any more, since 
the one measured at the lower height develops a fatter tail at low $T$ values, indicating the asymmetry in the mixing process evolution.}
\label{fig:pdfTz.DT04}
\end{center}
\end{figure}
In figures (\ref{fig:pdfTz.DT005}) and  (\ref{fig:pdfTz.DT04}) we plot the PDFs, for such heights, for $At=0.05$ and
$At=0.4$ respectively. In bothe cases, of course, the PDF corresponding to the upper height shows a peak at lower 
values of $T$ (close to the unmixed cold fluid value), and vice versa for the lower height. However, while for small $At$ the
two PDFs are symmetric with respect to the average temperature (in some sense they transform into each other upon reversal
around $T=1$), for the compressible case figure (\ref{fig:pdfTz.DT04}) displays a clear asymmetry, where the PDF measured at the 
lower $z-$location develops a more intense tail at low $T$ values, indicating that falling cold fluid spikes are faster (and mix slowly) than 
rising hot fluid bubbles. 

\section{Evolution of averaged quantities} \label{sec:conservation}

If we integrate equation (\ref{eq:H2}), multiplied bu $u_i$, over the whole volume, 
since the boundary conditions are periodic at
the vertical walls and set zero velocity (no-slip) at top and bottom plates, we obtain the following equation
for the mean kinetic energy:
\be \label{eq:energy}
\partial_t \langle \frac{\rho u^2}{2} \rangle = \langle \rho  g  u_z \rangle - \epsilon_{diss}
\ee
where $\langle \dots \rangle = 1/(L_x L_z) \int (\dots) dx dz$ denotes the space average, and 
$\epsilon_{diss} =\nu  \langle  (\partial_j u_i)^2 \rangle $. Equation (\ref{eq:energy}) indicates that the total
forcing, due to the gravitational field, is consumed partly being transformed in kinetic energy and partly
by dissipation. 
\begin{figure}
\begin{center}
\advance\leftskip-0.55cm
\includegraphics[width=0.5\textwidth]{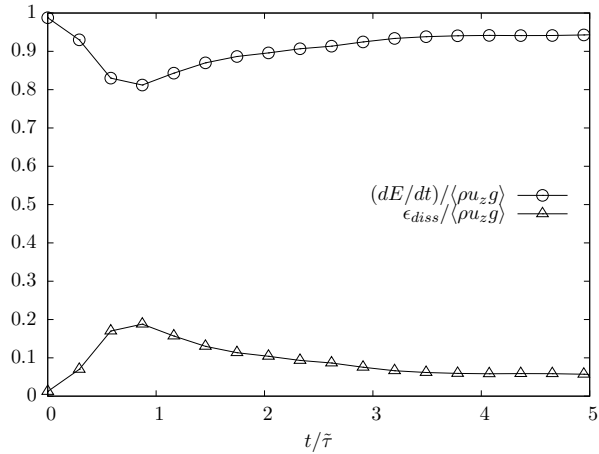}
\caption{Time derivative of the kinetic energy ($\odot$) and dissipation ($\triangle$) as functions of time, shown as fraction of 
the forcing term $\langle \rho g u_z \rangle$. Notice that the forcing due to gravity is almost entirely converted in kinetic energy, 
since the dissipation is basically negligible in 2D turbulence.}
\label{fig:energy2.DT005}
\end{center}
\end{figure}
In figure (\ref{fig:energy2.DT005}) we show the fraction of forcing transferred as kinetic energy 
($(dE/dt)/\langle \rho g u_z\rangle$) and dissipated ($\epsilon_{diss}/\langle \rho g u_z\rangle$), and we observe that 
the latter is much smaller than the former, as one would have expected, the dissipation being negligible in 2D. This behaviour is in strong contrast with what happens in 3D, where energy is transferred downscales and a sort of equipartition between kinetic energy growth and energy dissipation is achieved \cite{chertkovPOF2009,bof2010}.

\section{Conclusions} \label{sec:conclusions}
We simulated, by means of a new lattice Boltzmann algorithm, the turbulent dynamics of 
a Rayleigh-Taylor system, the characteristics of the method letting us tune the 
effects of both stratifications and compressibility. 
Concerning the former problem, we discussed the importance of the adiabatic gradient for the growth of
the RT mixing layer, showing the existence of the phenomenon of the arrest of the mixing process and 
of a maximal width, the adiabatic length, $L_{ad}$, for the mixing region. 
We measured, then, the PDFs of density and temperature fluctuations, inside the mixing region, observing that
while the two statistics are almost identical for small Atwood numbers (negligible compressibility),
they decouple when compressibility is large, owing to the increased relevance of pressure fluctuations, whose
PDF we also measured, confirming that for large temperature jumps pressure plays an active dynamical role.

\section{Acknowledgements} \label{sec:acknowledgements}

We acknowledge the collaboration of R. Tripiccione and F. Mantovani, for useful discussions and technical support.
One of the authors, AS, warmly thanks the TMB-2009 committee for the assignment of the ``young scientist award'' of the conference for the work
he presented on Eulerian and Lagrangian statistics in incompressible and weakly compressible turbulence. The choice of submitting, then, this new
material on Rayleigh-Taylor turbulence is due to the fact that it is more recent and, we think, of possibly greater interest for the TMB community.


\end{document}